# High Spectral Resolution Observations of Propynal (HCCCHO) towards TMC-1 from the GOTHAM Large Program on the Green Bank Telescope

Anthony J. Remijan,[1] Zachary T. P. Fried,[2] Ilsa R. Cooke,[3] Gabi Wenzel,[2] Ryan Loomis,[1] Christopher N. Shingledecker,[4] Andrew Lipnicky,[1] Ci Xue,[2] Michael C. McCarthy,[5] and Brett A. McGuire[2,1]

[1]*National Radio Astronomy Observatory, Charlottesville, VA 22903, USA*
[2]*Department of Chemistry, Massachusetts Institute of Technology, Cambridge, MA 02139, USA*
[3]*Department of Chemistry, University of British Columbia, 2036 Main Mall, Vancouver BC V6T 1Z1, Canada*
[4]*Department of Physics and Astronomy, Benedictine College, Atchison, KS 66002 USA*
[5]*Center for Astrophysics | Harvard & Smithsonian, Cambridge, MA 02138, USA*

## ABSTRACT

We used new high spectral resolution observations of propynal (HCCCHO) towards TMC-1 and in the laboratory to update the spectral line catalog available for transitions of HCCCHO — specifically at frequencies lower than 30 GHz which were previously discrepant in a publicly available catalog. The observed astronomical frequencies provided high enough spectral resolution that, when combined with high-resolution ($\sim$ 2 kHz) measurements taken in the laboratory, a new, consistent fit to both the laboratory and astronomical data was achieved. Now with a nearly exact (< 1 kHz) frequency match to the $J = 2 - 1$ and $3 - 2$ transitions in the astronomical data, using a Markov chain Monte Carlo (MCMC) analysis, a best fit to the total HCCCHO column density of $7.28^{+4.08}_{-1.94} \times 10^{12}$ cm$^{-2}$ was found with a surprisingly low excitation temperature of just over 3 K. This column density is around a factor of 5 times larger than reported in previous studies. Finally, this work highlights that care is needed when using publicly available spectral catalogs to characterize astronomical spectra. The availability of these catalogs is essential to the success of modern astronomical facilities and will only become more important as the next generation of facilities come online.

## 1. INTRODUCTION

One of the fundamental foundations in astrochemical studies, specifically in the detection of a new molecule (Snyder et al. 2005), is the accuracy of the calculated frequencies of molecular rotational transitions. As instrumentation on new astronomical facilities improves, it tests the limits to laboratory measurements. Laboratory measurements are typically the landmark for determining the millimeter and submillimeter wave spectra of molecules. The more accurate laboratory measurements are, and the broader the range of measurements across the radio and (sub)millimeter spectrum, the better the predictions are in guiding astronomical searches. The astronomical community routinely uses publicly available spectral line catalogs when searching for new molecules or for searching for transitions of well-known astronomical species. The state-of-the-art catalogs typically used for these investigations include the Jet Propulsion Laboratory (JPL)[1] (Pickett et al. 1998) and Cologne Database for Molecular Spectroscopy (CDMS)[2] (Endres et al. 2016) spectroscopic databases. For more than thirty years, these databases have set the standard for astronomical spectroscopy and the identification of transitions in a variety of astronomical sources. Over the last several years, new databases have come online including the Toyama Microwave Atlas for spectroscopists and astronomers[3], the Lille Spectroscopic Database[4], and the Splatalogue database for astronomical spectroscopy[5] that are also used for the identification of astronomical transitions.

---

[1] https://spec.jpl.nasa.gov/
[2] https://cdms.astro.uni-koeln.de/

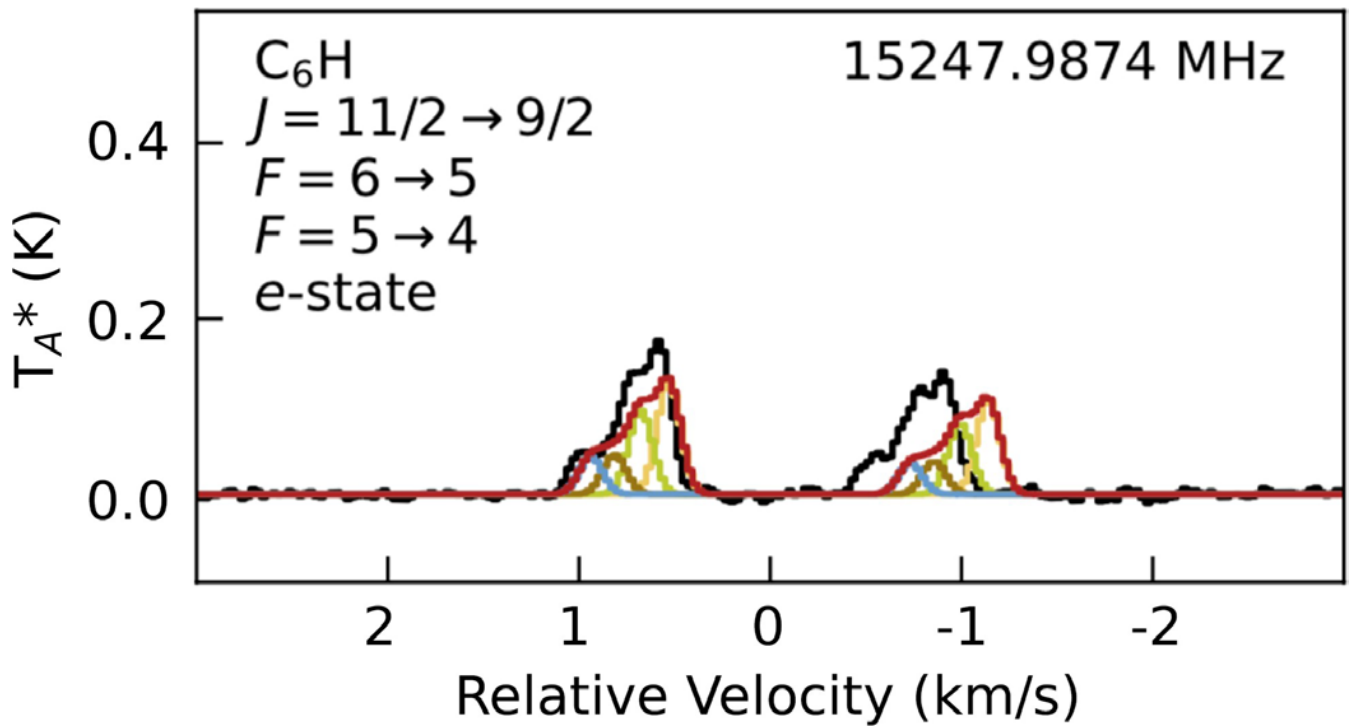

**Figure 1.** Model spectrum (red trace) of $C_6H$ based on transition rest frequencies taken from the CDMS online catalog (ID#:c073501 before the Dec 2023 update) compared to the astronomical spectrum (black trace) of TMC-1 around 15247.98 MHz. New fits to the $C_6H$ rest frequencies were reported in Remijan et al. (2023) and are available at https://cdms.astro.uni-koeln.de/ species ID#:c073501.

The recent detections of new astronomical molecules such as $C_{10}H^-$ (Remijan et al. 2023), $C_7N^-$ (Cernicharo et al. 2023a), HMgCCCN (Cabezas et al. 2023), $MgC_4H^+$, $MgC_3N^+$, $MgC_6H^+$, and $MgC_5N^+$ (Cernicharo et al. 2023b) show that in some special cases, the identification of new molecules can take place without their confirmation in the laboratory; however, these are very rare cases and laboratory spectra are typically a crucial prerequisite when characterizing astronomical spectra. In the process of characterizing the $C_{10}H^-$ spectrum from the astronomical data, Remijan et al. (2023) found that several of the rest frequencies of the smaller polyynes, namely $C_4H$ and $C_6H$, were not accurate enough to match the observational data taken from TMC-1. Figure 1 illustrates the issues found when trying to match the rest frequencies taken from the online catalog from CDMS (ID#:c073501 before the Dec 2023 update) compared to the astronomical spectrum. Based on the data collected by the 100 m Robert C. Byrd Green Bank Telescope (GBT), new rest frequencies were determined based on the astronomical observations and new catalogs were generated and made available with the $C_{10}H^-$ publication (see supplemental material available in Remijan et al. (2023) and updated catalog file available at CDMS).

In the process of identifying other molecular transitions in the data collected from TMC-1, a similar issue was found with the spectral features of propynal. Propynal (HCCCHO) was first identified by Irvine et al. (1988) towards TMC-1 using the 140-ft telescope of the National Radio Astronomy Observatory (NRAO) in Green Bank at 18650.3 MHz and the 45m telescope of the Nobeyama Radio Observatory (NRO) at 37290.1 MHz. The spectral resolution of these observations were 10 kHz (smoothed to 20 kHz) and 37 kHz, respectively. Since this initial detection, HCCCHO has been detected toward several other cold dark clouds (Loison et al. 2016) and in the extended envelopes surrounding Sgr B2 (Hollis et al. 2004; Requena-Torres et al. 2008) but so far has eluded detection towards high and low-mass hot molecular cores (Manigand et al. 2021). Recently, Cernicharo et al. (2021) observed 6 transitions of HCCCHO as part of the Q-band Ultrasensitive Inspection Journey to the Obscure Tmc-1 Environment (QUIJOTE) survey. The HCCCHO data were taken at both Ka and Q-band and shown between 37 – 46 GHz (see Figure D.1 of Cernicharo et al. (2021)). In Figure D.1, the 5(0,5)-4(0,4) simulation appears slightly offset from the observational data however, given the spectral resolution of the data, no definitive offset was determined. As such, from these observations, the observed spectral features match the simulated spectrum very well at a spectral resolution of 38.15 kHz. The authors derived a column density of $1.5 \times 10^{12}$ cm$^{-2}$ which is *identical* to the column density determined by (Irvine et al. 1988) from the initial detection.

---

[3] http://moodle.sci.u-toyama.ac.jp/atlas/

[4] https://lsd.univ-lille.fr/

[5] https://splatalogue.online/

In this work, we reinvestigate the original detection of the 18650.3 MHz transitions of HCCCHO ($K_a = 0$) presented in Irvine et al. (1988) and report the detection of the $K_a = 1$ features from this $J = 2 - 1$ transition series. We also show the discrepancy between the observed astronomical line frequencies toward TMC-1 taken with 1.4 kHz spectral resolution with the GBT Observations of TMC-1: Hunting Aromatic Molecules (GOTHAM) survey and from the laboratory measured line frequencies reported in catalog ID#: c054510 from CDMS. As such, we refit the HCCCHO spectral lines using new laboratory data and the spectral lines reported in Jabri et al. (2020) and Robertson et al. (2023). In Section 2 we describe the astronomical observations; Section 3 details the line fitting process of HCCCHO including new laboratory data; Section 4 describes the results of the new line fitting applied to the astronomical data; finally Section 5 discusses the discrepancies found between the newly collected data and the previously identified lines in this source and also the care that is needed in using online catalogs in the analysis of spectral line observations especially in light of new broadband, high spectral resolution, high sensitivity instrumentation. Section 6 summarizes our conclusions.

## 2. OBSERVATIONS

Observations for this study were obtained as part of the GOTHAM survey, a large program that utilized the 100-m GBT currently managed by the Green Bank Observatory (GBO). All data were collected with a uniform frequency resolution of 1.4 kHz (0.05–0.01 km/s in velocity) and a root-mean-square (RMS) noise of ∼2–20 mK across most of the observed frequency range, with the RMS noise gradually increasing toward higher frequencies due to shorter integration times.

This work employs the fourth data reduction (DR4) of GOTHAM, targeting the cyanopolyyne peak (CP) of TMC-1, centered at $\alpha_{J2000} = 04^\text{h} 41^\text{m} 42.5^\text{s}$, $\delta_{J2000} = +25°41'26.8''$. A comprehensive description of the fourth data reduction can be found in Sita et al. (2022), and the observing strategy and reduction pipeline are fully described in McGuire et al. (2020). Briefly, the spectra of the GOTHAM survey encompass the entirety of the X-, K-, and Ka-receiver bands with nearly continuous coverage from 7.9 to 11.6 GHz, 12.7 to 15.6 GHz, and 18.0 to 36.4 GHz (24.9 GHz of total bandwidth). The HCCCHO lines used in this analysis are limited specifically to K-band and include the $J = 2 - 1$ and $3 - 2$ series of transitions. Data reduction involved the removal of radio frequency interference (RFI) and artifacts, baseline continuum fitting, and flux calibration using complementary Very Large Array (VLA) observations of the source J0530+1331. The uncertainty from this flux calibration is estimated at approximately 20% and is factored into our statistical analysis described below.

## 3. SPECTROSCOPIC ANALYSIS

Figure 2 shows the astronomical data and the subsequent fits for four velocity components of HCCCHO that illustrates the issue present in the rest frequencies of the transitions taken from the CDMS catalog entry ID#054510 (red trace) and the measured frequencies from the GBT (black trace)[6]. A non-systematic offset was clearly detected in the frequencies, especially in the $J = 3 - 2$ transitions of HCCCHO (panels 2d-f). The microwave transitions contained within the online catalog were measured in a spectroscopic study from Costain & Morton (1959). As such, in order to determine more accurate rest frequencies of these lower-$J$ transitions, we refit the HCCCHO rotational spectrum using a combination of new laboratory data and more updated data present in the literature (Jabri et al. 2020; Robertson et al. 2023).

---

[6] The frequencies from the GBT are the measured sky frequencies shifted by the $v_{LSR}$ 5.8 km s$^{-1}$ of TMC-1. The velocity shifts of each individual source component are listed in section 4

**Table 1.** Newly measured HCCCHO rotational transitions using a Fabry-Perot cavity-based FTMW spectrometer at the Harvard–Smithsonian Center for Astrophysics (CfA). All measured frequencies have uncertainties of 2 kHz. The frequencies of these transitions determined by our new fit as well as the fit from Robertson et al. (2023) are also displayed. The listed differences are the difference in frequency between the measured frequency and the frequencies determined by the respective fits.

| Quantum Numbers $J' K_a' K_c' – J'' K_a'' K_c''$ | Measured Frequency (MHz) | Frequency from New Fit (MHz) | Difference (MHz) | Robertson et al. (2023) (MHz) | Difference (MHz) |
|---|---|---|---|---|---|
| 1 0 1 – 0 0 0 | 9325.8083 | 9325.8081 | 0.0002 | 9325.8079 | 0.0004 |
| 4 1 4 – 5 0 5 | 15146.0405 | 15146.0410 | -0.0005 | 15146.0434 | -0.0029 |
| 2 1 2 – 1 1 1 | 18325.5418 | 18325.5413 | 0.0005 | 18325.5408 | 0.001 |
| 2 0 2 – 1 0 1 | 18650.3084 | 18650.3080 | 0.0004 | 18650.3075 | 0.0009 |
| 2 1 1 – 1 1 0 | 18978.7837 | 18978.7830 | 0.0007 | 18978.7825 | 0.0012 |
| 3 1 3 – 4 0 4 | 25100.6616 | 25100.6617 | -0.0001 | 25100.6639 | -0.0023 |
| 9 0 9 – 8 1 8 | 26074.6867 | 26074.6838 | 0.0029 | 26074.6805 | 0.0062 |
| 3 1 3 – 2 1 2 | 27487.4323 | 27487.4318 | 0.0005 | 27487.4310 | 0.0013 |
| 3 0 3 – 2 0 2 | 27972.1934 | 27972.1921 | 0.0013 | 27972.1914 | 0.002 |
| 3 2 2 – 2 2 1 | 27980.7936 | 27980.7911 | 0.0025 | 27980.7904 | 0.0032 |
| 3 2 1 – 2 2 0 | 27985.8382 | 27985.8396 | -0.0014 | 27985.8388 | -0.0006 |
| 3 1 2 – 2 1 1 | 28467.2504 | 28467.2496 | 0.0008 | 28467.2489 | 0.0015 |
| 2 1 2 – 3 0 3 | 34903.3872 | 34903.3842 | 0.0030 | 34903.3860 | 0.0012 |
| 4 1 4 – 3 1 3 | 36648.2700 | 36648.2705 | -0.0005 | 36648.2695 | 0.0005 |
| 4 0 4 – 3 0 3 | 37290.1552 | 37290.1542 | 0.001 | 37290.1532 | 0.002 |
| 4 1 3 – 3 1 2 | 37954.6035 | 37954.6026 | 0.0009 | 37954.6016 | 0.0019 |

There have been several previous rotational studies of propynal (Costain & Morton 1959; Winnewisser 1973; Jaman et al. 2011; Barros et al. 2015; Jabri et al. 2020; Robertson et al. 2023). While the more recent studies (e.g., Jabri et al. (2020); Robertson et al. (2023)) extended the rotational spectrum into the sub-mm wave frequency regime, the measured microwave frequencies used in their global fits were in many cases from Costain & Morton (1959), which, as mentioned previously, have some notable frequency offsets from our astronomical observations. Thus, in order to determine more accurate rest frequencies of these lower-$J$ transitions (allowing for an improved astronomical analysis toward TMC-1), we remeasured these lines and refitted the HCCCHO rotational spectrum. The data used in our fit is a combination of three studies. The majority of the lines in the updated fit were measured by Robertson et al. (2023). This group measured and analyzed the absorption spectrum of HCCCHO from around 150–900 GHz. However, in their combined fit, the measured transitions for all lines below ∼150 GHz were from older rotational studies of propynal (Costain & Morton 1959; Winnewisser 1973; Jaman et al. 2011). In order to include more updated data below 150 GHz, our new fit includes the lines measured by Jabri et al. (2020). This group predominately measured lines from 82–480 GHz using a direct-absorption millimeter-wave spectrometer at room temperature. They also measured transitions from 6-18 GHz using a chirped-pulse Fourier transform microwave (FTMW) spectrometer. Thus, we included the transitions measured by Jabri et al. (2020) from around 82 - 150 GHz, along with a small number of microwave transitions. Finally, in order to measure even more precise and accurate rotational frequencies of the transitions within our GOTHAM TMC-1 observations, we have remeasured several microwave transitions ranging from ∼9–38 GHz using a Fabry-Perot cavity-based FTMW spectrometer at the Harvard–Smithsonian Center for Astrophysics (CfA). These lines are estimated to have uncertainties of 2 kHz, whereas the microwave transitions previously measured by Costain & Morton (1959) had estimated uncertainties of 20 kHz. The newly measured lines are listed in Table 1

Overall, the updated fit contains 2674 unique transition frequencies. Including blended lines, this corresponds to 3230 transitions. Of these distinct frequencies, 16 were measured at the CfA, 178 were measured by Jabri et al. (2020), and 2480 were measured by Robertson et al. (2023). The transitions in the fit range from $J'' = 1$–100 and $K_a'' = 0$–15. Of these transitions, 1493 are $\mu_a$ transitions and 1737 are $\mu_b$ transitions. There are no c-type transitions because HCCCHO is a planar molecule that has no permanent dipole moment along its c axis. This combined dataset was analyzed with a least-squares fitting procedure using SPFIT in Pickett's CALPGM suite of programs (Pickett 1991). As was done by Robertson et al. (2023), the standard errors were then computed using PIFORM. The fit was done using the S-reduction with the $I^r$ representation. The updated fit includes 27 rotational and distortion constants, as presented in Table 2. By updating the lower frequency data, the new fit includes a greater number of transitions and has an improved RMS error while requiring two fewer distortion parameters to accurately fit the spectrum. The uncertainty of each parameter has also decreased in the new fit. Additionally, as can be seen in Table 1, the updated fit very precisely matches the newly measured microwave lines in the frequency range of the GOTHAM TMC-1 observations. In fact, compared to the previous parameter set, the updated fit has the same or smaller deviations for 14 of the 16 newly measured lines. The full measured line list, including which lines are used from which data sources, along with the corresponding input and output files from SPFIT/SPCAT, are provided as Supplemental Information.

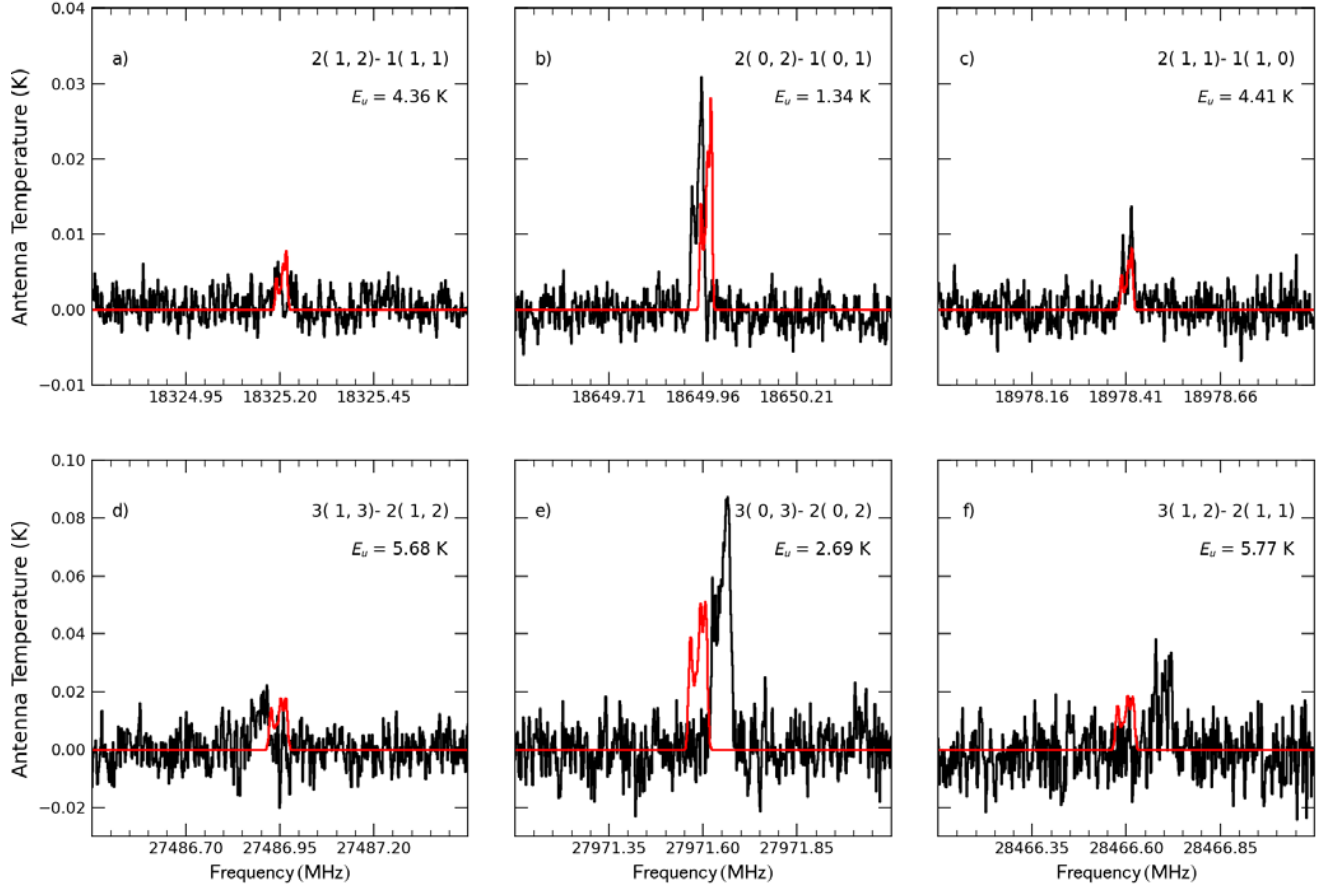

**Figure 2.** Astronomical data taken as part of the GOTHAM program and the subsequent fits for four velocity components of HCCCHO, illustrating the mismatch in the rest frequencies taken from the CDMS catalog (red trace) and the measured frequencies from the GBT (black trace). HCCCHO transition quantum numbers and upper state energy levels (K) are given in the top right corner of each panel.

## 4. ASTRONOMICAL ANALYSIS

The best-fit physical parameters including column density [$N_T$], excitation temperature [$T_{ex}$], line width [$\Delta V$], $v_{LSR}$ and source size (″) were determined using the Markov chain Monte Carlo (MCMC) model as described in previous GOTHAM analyses (see, e.g., Remijan et al. 2023; Sita et al. 2022; Siebert et al. 2022; Lee et al. 2021) and discussed in detail in Loomis et al. (2021). In short, the MCMC model calculates the probability distributions and covariances for the parameters used to describe the emission of molecules observed in our data over four distinct velocity components. This analysis takes into account a background continuum temperature of 2.7 K resulting from the cosmic microwave background. 86 catalog transitions within the frequency range of the GOTHAM TMC-1 observations were considered in the MCMC analysis, although only 22 of these are simulated to have any appreciable intensity above or below the noise. Our MCMC ran for 50,000 iterations with 100 walkers. Uniform priors were utilized for the column densities, excitation temperatures, and source sizes. However, more strict Gaussian priors were used on the $v_{LSR}$ values. These were centered on the known source velocities from previous molecular studies. Additionally, in order for the MCMC model to sufficiently converge, we needed to restrict the linewidth to values around 0.12 km/s, which is similar to the linewidth of several other organic molecules in this source (e.g. Xue et al. 2020; Cooke et al. 2023; Lee et al. 2021; Shingledecker et al. 2021). The resulting corner plot from the MCMC analysis of HCCCHO is shown in Figure A1. As in our previous analyses, we adopted the 50th percentile value of the posterior probability distributions as the representative value of each parameter. We then use the 16th and 84th percentile values for the uncertainties, corresponding to $\pm 1\sigma$ for a Gaussian posterior distribution. Table 3 lists the source parameters determined for each of our four velocity components of HCCCHO. From this analysis, we find a total HCCCHO column density of $7.28^{+4.08}_{-1.94} \times 10^{12}$ cm$^{-2}$ at an excitation temperature of just over 3 K. Even though the measured excitation temperature is quite low, this temperature value is sufficient to explain the measured intensities of (nearly) all transitions detected from our observations including the $K_a$=1 components for both the $J = 2 - 1$ and $3 - 2$ transitions. This low temperature may also explain the lack of any detection of the $K_a$=2 components of the $J = 3 - 2$ transition, which have upper state energies > 14 K.

**Table 2.** Rotational parameters of HCCCHO from this work along with previous fits.

| Parameter | Current Work | Robertson et al. (2023) | CDMS Constants |
|---|---|---|---|
| $A$ (MHz) | 68035.25798(32) | 68035.25904(49) | 68035.3 |
| $B$ (MHz) | 4826.223741(22) | 4826.223611(40) | 4826.22 |
| $C$ (MHz) | 4499.591856(23) | 4499.591734(40) | 4499.59 |
| | | | |
| $D_J$ (MHz) | $1.871682(11) \times 10^{-3}$ | $1.871637(17) \times 10^{-3}$ | |
| $D_{JK}$ (MHz) | $-0.14772431(48)$ | $-0.14772496(65)$ | |
| $D_K$ (MHz) | $8.990458(11)$ | $8.990558(16)$ | |
| $d_1$ (MHz) | $-3.460695(24) \times 10^{-4}$ | $-3.460680(32) \times 10^{-4}$ | |
| $d_2$ (MHz) | $-2.101133(86) \times 10^{-5}$ | $-2.10110(11) \times 10^{-5}$ | |
| | | | |
| $H_J$ (MHz) | $5.9747(20) \times 10^{-9}$ | $5.9684(29) \times 10^{-9}$ | |
| $H_{JK}$ (MHz) | $-7.6935(21) \times 10^{-7}$ | $-7.6954(27) \times 10^{-7}$ | |
| $H_{KJ}$ (MHz) | $-6.5558(75) \times 10^{-6}$ | $-6.5603(94) \times 10^{-6}$ | |
| $H_K$ (MHz) | $2.40856(14) \times 10^{-3}$ | $2.41053(17) \times 10^{-3}$ | |
| $h_1$ (MHz) | $2.24082(56) \times 10^{-9}$ | $2.24051(74) \times 10^{-9}$ | |
| $h_2$ (MHz) | $2.9049(31) \times 10^{-10}$ | $2.9039(39) \times 10^{-10}$ | |
| $h_3$ (MHz) | $8.4014(74) \times 10^{-11}$ | $8.3998(99) \times 10^{-11}$ | |
| | | | |
| $L_J$ (MHz) | $-3.186(12) \times 10^{-14}$ | $-3.157(16) \times 10^{-14}$ | |
| $L_{JJK}$ (MHz) | $6.328(36) \times 10^{-12}$ | $6.341(46) \times 10^{-12}$ | |
| $L_{JK}$ (MHz) | $-4.164(15) \times 10^{-10}$ | $-4.146(20) \times 10^{-10}$ | |
| $L_{KKJ}$ (MHz) | $-4.455(40) \times 10^{-9}$ | $-4.493(52) \times 10^{-9}$ | |
| $L_K$ (MHz) | $-5.0299(66) \times 10^{-7}$ | $-5.1634(76) \times 10^{-7}$ | |
| $l_1$ (MHz) | $-1.4139(39) \times 10^{-14}$ | $-1.4119(51) \times 10^{-14}$ | |
| $l_2$ (MHz) | $-2.573(28) \times 10^{-15}$ | $-2.564(35) \times 10^{-15}$ | |
| $l_3$ (MHz) | $-1.589(13) \times 10^{-15}$ | $-1.588(16) \times 10^{-15}$ | |
| $l_4$ (MHz) | $-2.866(24) \times 10^{-16}$ | $-2.876(31) \times 10^{-16}$ | |
| | | | |
| $P_{JK}$ (MHz) | $-6.54(21) \times 10^{-17}$ | $-6.47(27) \times 10^{-17}$ | |
| $P_{KJ}$ (MHz) | $5.00(11) \times 10^{-13}$ | $4.71(14) \times 10^{-13}$ | |
| $P_{KKJ}$ (MHz) | $-1.3804(64) \times 10^{-12}$ | $-1.3697(86) \times 10^{-12}$ | |
| $P_K$ (MHz) | – | $2.68(12) \times 10^{-11}$ | |
| | | | |
| $T_K$ (MHz) | – | $1.801(56) \times 10^{-14}$ | |
| | | | |
| Transitions in fit | 3230 | 3111 | |
| $\sigma_{\text{fit}}$ (MHz) | 0.038 | 0.04 | |

To illustrate how well our derived parameters using our MCMC approach can reproduce the astronomically measured spectrum, a model of the molecular emission is generated for each set of parameters (i.e. the column densities, excitation temperatures, line widths, $v_{LSR}$, and source sizes determined for each velocity component). The simulation was conducted using the `molsim` software package (McGuire et al. 2024) following the conventions of Turner (1991) for a single excitation temperature (in this case, for all four velocity components) and accounting for the effect of optical depth. Prior observations from GOTHAM (Xue et al. 2020) and others (Dobashi et al. 2018, 2019) have found that most emission seen at centimeter wavelengths in TMC-1 can be separated into contributions from four distinct velocity components within the larger structure. The exact velocity values of these four velocity components as determined by our MCMC analysis in this work is presented in Table 3. Figure 3 shows the results of the new catalog fit (orange trace) overlaid on the measured frequencies from the GBT (black trace). As Figure 3 illustrates, the rest frequencies of each of the transitions now perfectly match (< 1 kHz) the observed astronomical frequencies and the derived intensity values determined from the MCMC model parameters for each transition are also in good agreement with the observed astronomical intensities. For the $K_a$=1 transitions, the intensities are well within the 1$\sigma$ RMS values. However, the predicted intensity of the 3(0,3)-2(0,2) transition only matches the observed astronomical intensity at about the 60% level. This discrepancy will be discussed further in the following section.

**Table 3.** Summary statistics of the marginalized HCCCHO posterior.

| $v_{lsr}$ (km s$^{-1}$) | Size (″) | $N_T$ ($10^{12}$cm$^{-2}$) | $T_{ex}$ (K) | $\Delta V$ (km s$^{-1}$) |
|---|---|---|---|---|
| $5.608^{+0.008}_{-0.009}$ | $60^{+121}_{-26}$ | $1.16^{+0.80}_{-0.41}$ | | |
| $5.750^{+0.012}_{-0.013}$ | $28^{+13}_{-8}$ | $1.89^{+1.38}_{-0.77}$ | $3.08^{+0.16}_{-0.12}$ | $0.120^{+0.000}_{-0.000}$ |
| $5.885^{+0.025}_{-0.024}$ | $20^{+17}_{-9}$ | $1.28^{+2.28}_{-0.77}$ | | |
| $6.034^{+0.010}_{-0.010}$ | $17^{+10}_{-5}$ | $2.95^{+2.97}_{-1.55}$ | | |

$N_T$(Total): $7.28^{+4.08}_{-1.94} \times 10^{12}$ cm$^{-2}$

**Table 4.** Molecular line parameters of HCCCHO including the differing line frequencies of the strongest expected HCCCHO transitions within the GOTHAM observations of TMC-1 using the new and previous spectroscopic catalogs.

| Quantum Numbers $J' K_a' K_c' - J'' K_a'' K_c''$ | This Work (MHz) | CDMS Catalog (MHz) | Difference (MHz) | Upper State Energy (K) | Line Strength (D$^2$) |
|---|---|---|---|---|---|
| 1 0 1 – 0 0 0 | 9325.8081(1) | 9325.8046(28) | 0.0035 | 0.45 | 5.56 |
| 2 1 2 – 1 1 1 | 18325.5413(2) | 18325.5600(1000) | -0.0187 | 4.36 | 8.35 |
| 2 0 2 – 1 0 1 | 18650.3080(2) | 18650.3300(1000) | -0.0220 | 1.34 | 11.13 |
| 2 1 1 – 1 1 0 | 18978.7830(2) | 18978.7800(1000) | 0.0030 | 4.41 | 8.35 |
| 3 1 3 – 2 1 2 | 27487.4318(3) | 27487.4800(1000) | -0.0482 | 5.68 | 14.84 |
| 3 0 3 – 2 0 2 | 27972.1921(3) | 27972.1300(1000) | 0.0621 | 2.69 | 16.70 |
| 3 1 2 – 2 1 1 | 28467.2496(3) | 28467.1500(1000) | 0.0996 | 5.77 | 14.84 |

NOTE – Frequency errors reported are 1$\sigma$ errors on the last set of significant figures in the observational rest frequency.

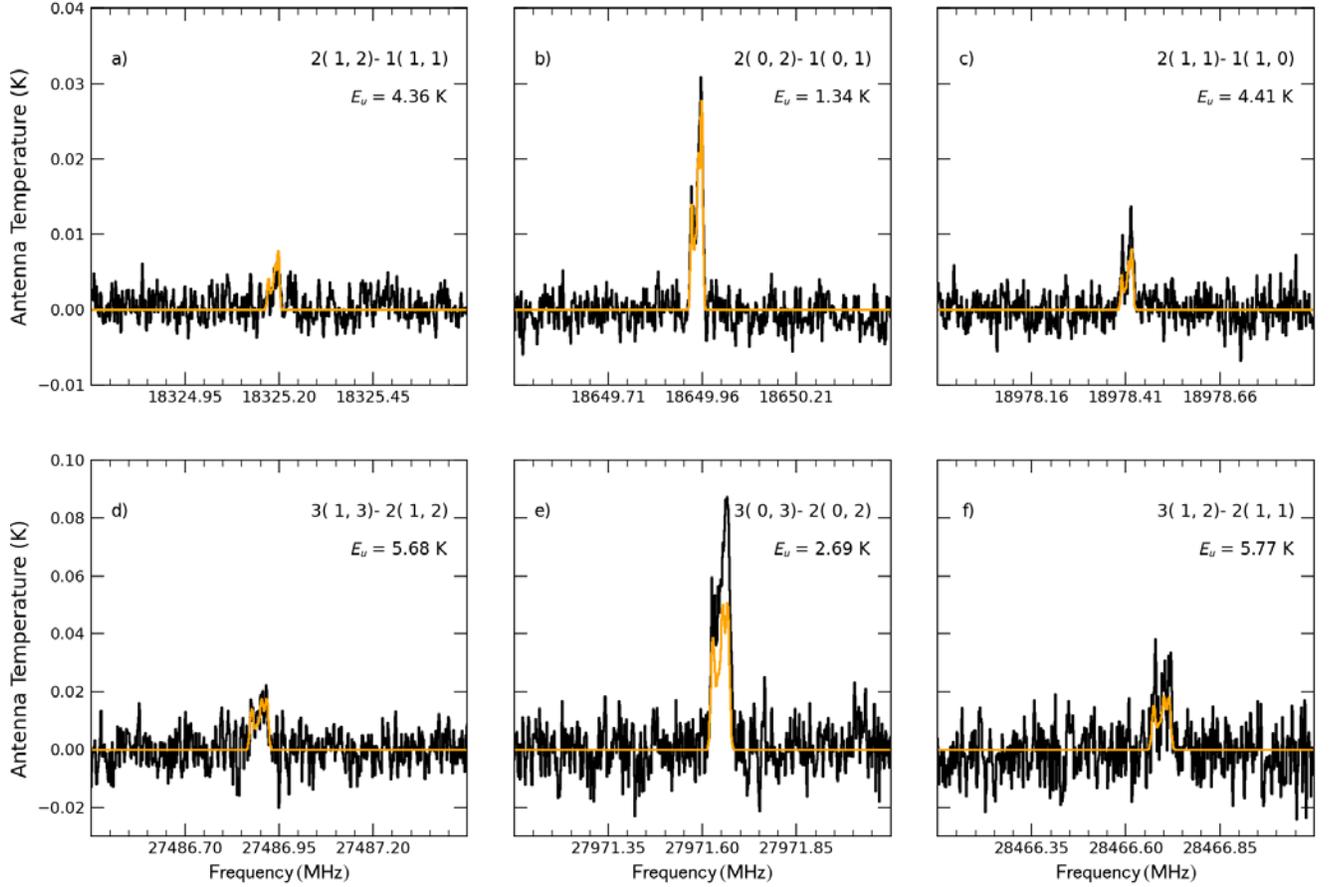

**Figure 3.** Astronomical data taken as part of the GOTHAM large program on the GBT and the subsequent fits for four velocity components of HCCCHO based on the new spectral line fitting (orange trace) and the measured frequencies from the GBT (black trace). Panel description is the same as in Figure 2.

## 5. DISCUSSION

The results presented in this work illustrate the need to fully characterize the spectrum of a molecule such as HCCCHO in determining the physical environment of a region — even one as "simple" as TMC-1. For example, the detection of the 4(0,4) - 3(0,3) transition at 37290.1 MHz from both the original detection paper of Irvine et al. (1988) and Cernicharo et al. (2021) show a factor of $\sim$ 2 discrepancy in the observed line intensity at nearly the same spectral resolution. Using nearly identical facilities, i.e. the NRO 45-m and the Yebes 40-m, the measured line intensities are 43 mK and $\sim$23 mK, respectively. Yet, there is no discussion of why this discrepancy exists between these two sets of observations. The predictions from this work based on the model parameters found from the MCMC analysis suggest an intensity for both the 4(0,4)-3(0,3) and 5(0,5)-4(0,4) at 46602.9 MHz, to be $\sim$ 80 mK observed with the GBT. Additional observing time for these transitions, as well as the $J = 1 - 0$ transition at 9325.8 MHz which is predicted at an intensity of $\sim$15 mK, will help to better characterize the physical environment of this source, especially since Loison et al. (2016) detected the 9(0,9)-8(0,8) transition of HCCCHO at 83775.8 MHz in TMC-1 with an upper state energy level >20 K.

A further difference is found in the determined column density of HCCCHO from the previous observations and this work. Given the low excitation temperature constrained by the limited energy level coverage of our detected transitions, a column density of $7.28^{+4.08}_{-1.94} \times 10^{12}$ cm$^{-2}$ was obtained which is a factor of $\sim$5 times the column density determined from Irvine et al. (1988) and Cernicharo et al. (2021) who reported a column density closer to $1.5 \times 10^{12}$ cm$^{-2}$ and $\sim$10$\times$ larger than Loison et al. (2016) who found a column density of $\sim 8 \times 10^{11}$ cm$^{-2}$. Some of this difference can be accounted for given the temperature used by each investigation. Irvine et al. (1988) and Loison et al. (2016) both assumed an excitation temperature of 10 K in determining the total HCCCHO column density. However,

Cernicharo et al. (2021) used 5 K and 4 K for the $K_a$=0 and $K_a$=1 transitions, respectively. To try and account for the column density discrepancy between these new observations on the GBT and the results found from Cernicharo et al. (2021) and Loison et al. (2016), we made the following assumptions: 1) using an excitation temperature of 5 K and; 2) assuming that the individual sources detected fill the GBT primary beam, we recalculated the column density and found a column density of $\sim 7.67 \times 10^{11}$ cm$^{-2}$ which is nearly identical to the column density found by Loison et al. (2016) and about a factor of 2 smaller than what Cernicharo et al. (2021) determined. As illustrated, the determination of the column densities (or abundances) of molecular species in astronomical regions are highly dependent on the measured (or assumed) physical conditions. And with our determined excitation temperature of $\sim$3 K and source sizes from this work, a high column density is needed given the measured intensity of the observed transitions. However, additional data are needed to test the robustness of both the temperature and source size determination (see below).

Upon examining the data closely, the MCMC-derived parameters do not fully predict the astronomically measured line intensity of the 3(0,3)-2(0,2) transition (see Figure 3e) - in fact, the column density of HCCCHO would need to be $\sim$ 40% larger to account for the discrepancy between the model spectrum and the data. Cernicharo et al. (2021) found that the $K_a$=0 and $K_a$=1 transitions of HCCCHO were best fit by 2 different excitation temperatures - namely 5 K and 4 K, respectively. While we do not directly determine the need to apply a different temperature between the $K_a$=0 and $K_a$=1 transitions, it is possible there may be anomalous absorption (or emission) from the low energy $K_a$=0 transitions due to the influence of the cosmic microwave background (CMB) or other sources. In order to pursue any further analysis, additional astronomically observed transitions are needed to better determine the physical environment (and specifically the non-LTE excitation properties) of this molecule. In addition, collision rates of HCCCHO–H$_2$ and HCCCHO–He are unknown so a full statistical equilibrium analysis is not possible and also beyond the scope of this paper. Now, it is well known that transitions of formaldehyde (H$_2$CO) are seen in absorption towards TMC-1 - these are the very low energy transitions absorbing against the CMB (Henkel et al. 1981; Townes & Cheung 1969). It is possible that given the structural similarity of HCCCHO and H$_2$CO, the lowest energy transitions of HCCCHO may show a similar effect which is why the detection of the $J = 1 - 0$ fundamental transition toward TMC-1 is critical in determining the physical environment. A more pragmatic explanation to this discrepancy however, is that the absolute amplitude calibration within this frequency range is incorrect. Previous GOTHAM observations have shown flux density errors on the order of $\sim$20% and measures are being taken to mitigate many of these inconsistencies in future data reductions. While a flux density error would account for some of the discrepancy, additional factors still need to be considered to account for the remainder. Hence, the observations of the higher $J$-value transitions at higher frequencies on the GBT with a robust flux density scale determination will be critical to resolving the anomalously high emission of this transition.

Finally, this work highlights the care that is needed when using publicly available catalogs to characterize astronomical spectra. The availability of these catalogs are absolutely critical to the success of modern astronomical facilities and will only become more important as the next generation of facilities come online — namely the next-generation Very Large Array (ngVLA) and the Atacama Large Millimeter/submillimeter Array Wideband Sensitivity Upgrade (WSU) (Carpenter et al. 2023). However, as this work and Remijan et al. (2023) have demonstrated, it is essential to understand the limitations of these catalogs. And without an understanding of both the source and molecule under investigation, significant effort is needed in order to verify a new claimed astronomical detection (see e.g., Dhariwal et al. 2024; Schuessler et al. 2022; Kolesniková et al. 2022; Snyder et al. 2005).

## 6. CONCLUSIONS

In summary, utilizing the high spectral resolution capabilities of the instrumentation on the GBT, we identified several frequency offsets in the low-$J$ transitions of HCCCHO toward the dark cloud source TMC-1. Using a combination of literature data and new laboratory data from the Harvard–Smithsonian CfA, an updated spectroscopic fit was conducted for HCCCHO, which was then used to compare to the astronomical data. Now with an exact frequency match to the $J = 2 - 1$ and $3 - 2$ transitions in the astronomical data, using an MCMC analysis, a best fit to the total HCCCHO column density of $7.28^{+4.08}_{-1.94} \times 10^{12}$ cm$^{-2}$ was found with a very low excitation temperature of just over 3 K. Nearly all of the measured intensities of the detected astronomical transitions are well matched using our determined model parameters but a $\sim$40% discrepancy is seen with the 3(0,3)-2(0,2) transition. It is possible this discrepancy may be due to absolute amplitude calibration uncertainties or possibly these low energy transitions are influenced by the CMB. Observations of the fundamental $J = 1 - 0$ transition as well as higher energy transitions are necessary to resolve this discrepancy. Finally, publicly available spectral line catalogs are absolutely critical in investigating the physical, chemical, kinematic, and dynamical conditions of astronomical sources; yet care must be taken when using these catalogs and understanding their limitations when used to identify transitions in astronomical sources.

## 7. DATA ACCESS & CODE

The code used to perform the analysis is part of the `molsim` open-source package ([McGuire et al. 2024](#)).

*Facilities:* GBT

We gratefully acknowledge support from NSF grants AST-1908576 and AST-2205126. G.W. and B.A.M. acknowledge the support of the Arnold and Mabel Beckman Foundation Beckman Young Investigator Award. ZTPF and BAM gratefully acknowledge the support of Schmidt Family Futures. I.R.C. acknowledges support from the University of British Columbia and the Natural Sciences and Engineering Research Council of Canada (NSERC). The National Radio Astronomy Observatory is a facility of the National Science Foundation operated under cooperative agreement by Associated Universities, Inc. The Green Bank Observatory is a facility of the National Science Foundation operated under cooperative agreement by Associated Universities, Inc.

## REFERENCES

Barros, J., Appadoo, D., McNaughton, D., et al. 2015, Journal of Molecular Spectroscopy, 307, 44, doi: 10.1016/j.jms.2014.12.011

Cabezas, C., Pardo, J. R., Agúndez, M., et al. 2023, A&A, 672, L12, doi: 10.1051/0004-6361/202346462

Carpenter, J., Brogan, C., Iono, D., & Mroczkowski, T. 2023, in Physics and Chemistry of Star Formation: The Dynamical ISM Across Time and Spatial Scales, ed. V. Ossenkopf-Okada, R. Schaaf, I. Breloy, & J. Stutzki, 304, doi: 10.48550/arXiv.2211.00195

Cernicharo, J., Cabezas, C., Endo, Y., et al. 2021, A&A, 650, L14, doi: 10.1051/0004-6361/202141297

Cernicharo, J., Pardo, J. R., Cabezas, C., et al. 2023a, A&A, 670, L19, doi: 10.1051/0004-6361/202245816

Cernicharo, J., Cabezas, C., Pardo, J. R., et al. 2023b, A&A, 672, L13, doi: 10.1051/0004-6361/202346467

Cooke, I. R., Xue, C., Changala, P. B., et al. 2023, The Astrophysical Journal, 948, 133, doi: 10.3847/1538-4357/acc584

Costain, C. C., & Morton, J. R. 1959, The Journal of Chemical Physics, 31, 389, doi: 10.1063/1.1730364

Dhariwal, A., Speak, T. H., Zeng, L., et al. 2024, ApJL, 968, L9, doi: 10.3847/2041-8213/ad4d9a

Dobashi, K., Shimoikura, T., Nakamura, F., et al. 2018, The Astrophysical Journal, 864, 82, doi: 10.3847/1538-4357/aad62f

Dobashi, K., Shimoikura, T., Ochiai, T., et al. 2019, The Astrophysical Journal, 879, 88, doi: 10.3847/1538-4357/ab25f0

Endres, C. P., Schlemmer, S., Schilke, P., Stutzki, J., & Müller, H. S. 2016, Journal of Molecular Spectroscopy, 327, 95, doi: https://doi.org/10.1016/j.jms.2016.03.005

Henkel, C., Wilson, T. L., & Pankonin, V. 1981, A&A, 99, 270

Hollis, J. M., Jewell, P. R., Lovas, F. J., Remijan, A., & Møllendal, H. 2004, ApJL, 610, L21, doi: 10.1086/423200

Irvine, W. M., Brown, R. D., Cragg, D. M., et al. 1988, ApJL, 335, L89, doi: 10.1086/185346

Jabri, A., Kolesniková, L., Alonso, E. R., et al. 2020, Journal of Molecular Spectroscopy, 372, 111333, doi: 10.1016/j.jms.2020.111333

Jaman, A. I., Bhattacharya, R., Mandal, D., & Das, A. K. 2011, Journal of Atomic and Molecular Physics, 2011, 439019, doi: 10.1155/2011/439019

Kolesniková, L., Belloche, A., Koucký, J., et al. 2022, A&A, 659, A111, doi: 10.1051/0004-6361/202142448

Lee, K. L. K., Loomis, R. A., Burkhardt, A. M., et al. 2021, The Astrophysical Journal Letters, 908, L11, doi: 10.3847/2041-8213/abdbb9

Loison, J.-C., Agúndez, M., Marcelino, N., et al. 2016, MNRAS, 456, 4101, doi: 10.1093/mnras/stv2866

Loomis, R. A., Burkhardt, A. M., Shingledecker, C. N., et al. 2021, Nature Astronomy, 5, 188, doi: 10.1038/s41550-020-01261-4

Manigand, S., Coutens, A., Loison, J. C., et al. 2021, A&A, 645, A53, doi: 10.1051/0004-6361/202038113

McGuire, B. A., Xue, C., Lee, K. L. K., El-Abd, S., & Loomis, R. A. 2024, molsim, v0.5.0, Zenodo, doi: 10.5281/zenodo.12697227

McGuire, B. A., Burkhardt, A. M., Loomis, R. A., et al. 2020, The Astrophysical Journal Letters, 900, L10, doi: 10.3847/2041-8213/aba632

Pickett, H. M. 1991, Journal of Molecular Spectroscopy, 148, 371, doi: 10.1016/0022-2852(91)90393-O

Pickett, H. M., Poynter, R. L., Cohen, E. A., et al. 1998, JQSRT, 60, 883, doi: 10.1016/S0022-4073(98)00091-0

Remijan, A., Scolati, H. N., Burkhardt, A. M., et al. 2023, ApJL, 944, L45, doi: 10.3847/2041-8213/acb648

Requena-Torres, M. A., Martín-Pintado, J., Martín, S., & Morris, M. R. 2008, ApJ, 672, 352, doi: 10.1086/523627

Robertson, E. G., Ruzi, M., McNaughton, D., et al. 2023, Journal of Molecular Spectroscopy, 394, 111786, doi: 10.1016/j.jms.2023.111786

Schuessler, C., Remijan, A., Xue, C., et al. 2022, ApJ, 941, 102, doi: 10.3847/1538-4357/ac8668

Shingledecker, C. N., Lee, K. L. K., Wandishin, J. T., et al. 2021, Astronomy & Astrophysics, 652, L12, doi: 10.1051/0004-6361/202140698

Siebert, M. A., Lee, K. L. K., Remijan, A. J., et al. 2022, ApJ, 924, 21, doi: 10.3847/1538-4357/ac3238

Sita, M. L., Changala, P. B., Xue, C., et al. 2022, ApJL, 938, L12, doi: 10.3847/2041-8213/ac92f4

Snyder, L. E., Lovas, F. J., Hollis, J. M., et al. 2005, ApJ, 619, 914, doi: 10.1086/426677

Townes, C. H., & Cheung, A. C. 1969, The Astrophysical Journal, 157, L103, doi: 10.1086/180395

Turner, B. E. 1991, The Astrophysical Journal Supplement Series, 76, 617, doi: 10.1086/191577

Winnewisser, G. 1973, Journal of Molecular Spectroscopy, 46, 16, doi: 10.1016/0022-2852(73)90023-4

Xue, C., Willis, E. R., Loomis, R. A., et al. 2020, The Astrophysical Journal Letters, 900, L9, doi: 10.3847/2041-8213/aba631

## APPENDIX

### A. HCCCHO ANALYSIS

The best-fit parameters from the MCMC fit for HCCCHO are shown in Table 3 and the priors used in determination of these parameters are given in Table A1. The corner plot from the analysis is shown in Figure A1. To measure the significance of detection, a spectral stack and matched filter analysis was then performed on the new HCCCHO spectral line catalog generated in this work. The resulting velocity stacked spectra and matched filter response of HCCCHO are shown in Figure A2. 22 low $K_a$ lines of propynal were included in the stack. We find a peak impulse response of $42.6\sigma$ from this analysis.

**Table A1.** Priors that were used for the MCMC analysis of the HCCCHO emission toward TMC-1. In this table, $U\{a, b\}$ indicates a uniform (unweighted) distribution between the two listed values. $N(\mu, \sigma^2)$ indicates a normal (Gaussian) distribution with mean, $\mu$, and variance, $\sigma^2$. Note that the listed values are simply for the final MCMC run that is presented in the paper. Letting the temperature prior range up to 10 K still resulted in convergence to low values within this range, so the prior was narrowed to reduce computational expense in the final run.

| Component No. | $v_{\rm LSR}$ [km s$^{-1}$] | Size ["] | $\log_{10}(N_T)$ [cm$^{-2}$] | $T_{\rm ex}$ [K] | $\Delta V$ [km s$^{-1}$] |
|---|---|---|---|---|---|
| 1 | $N(5.603, 0.005)$ | $U\{1, 300\}$ | | | |
| 2 | $N(5.747, 0.005)$ | $U\{1, 60\}$ | $U\{11.0, 13.5\}$ | $U\{2.7, 4.2\}$ | $N(0.120, 0.001)$ |
| 3 | $N(5.930, 0.005)$ | $U\{1, 70\}$ | | | |
| 4 | $N(6.036, 0.005)$ | $U\{1, 150\}$ | | | |

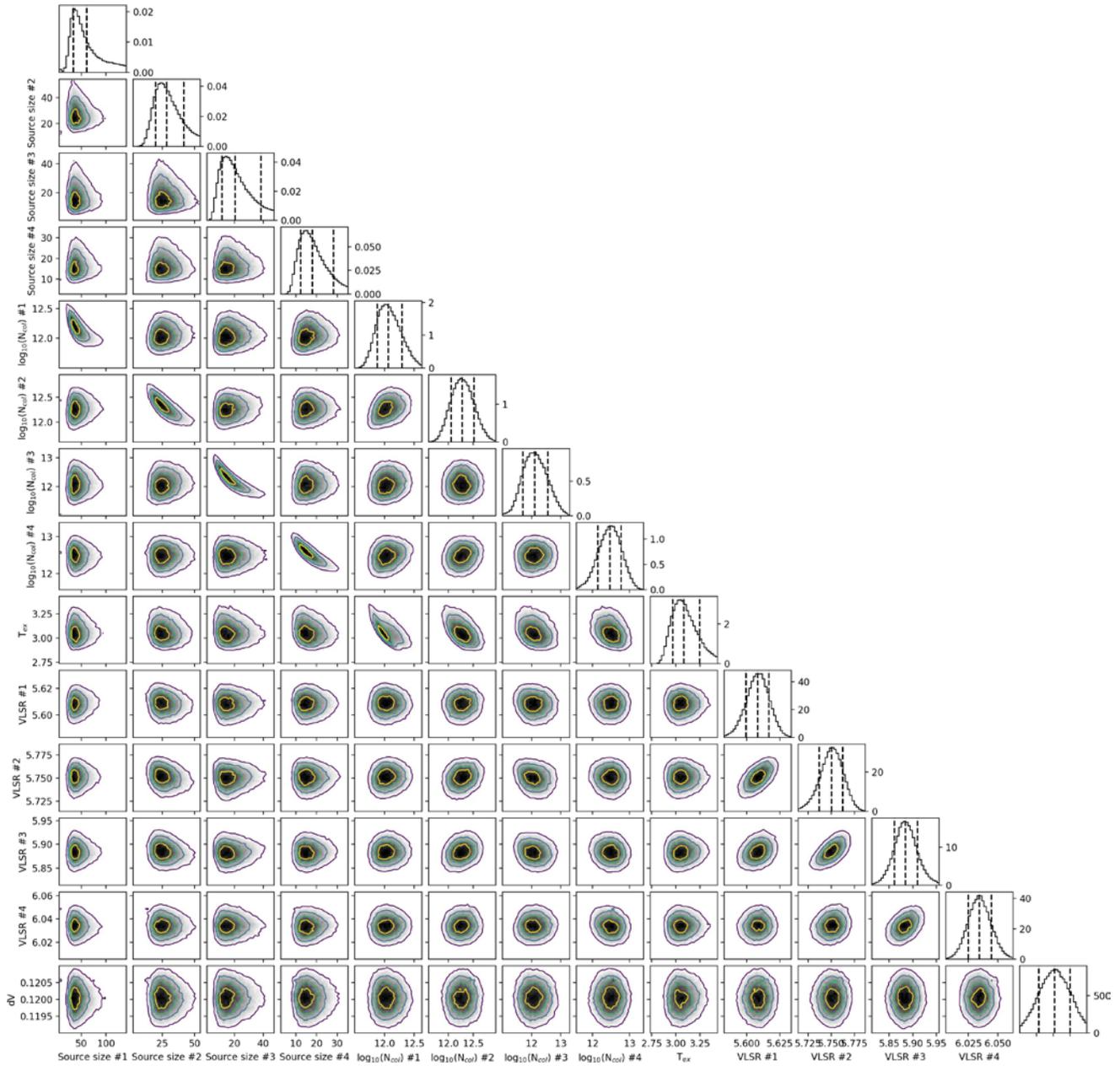

**Figure A1.** Corner plot generated from the MCMC analysis of the physical parameters of HCCCHO with 50,000 iterations. Covariance between the parameters is displayed in the off-diagonal panels while marginalized posterior distributions are shown in the panels on the diagonal. The vertical lines on the posterior distributions denote the 16th, 50th, and 84th confidence intervals.

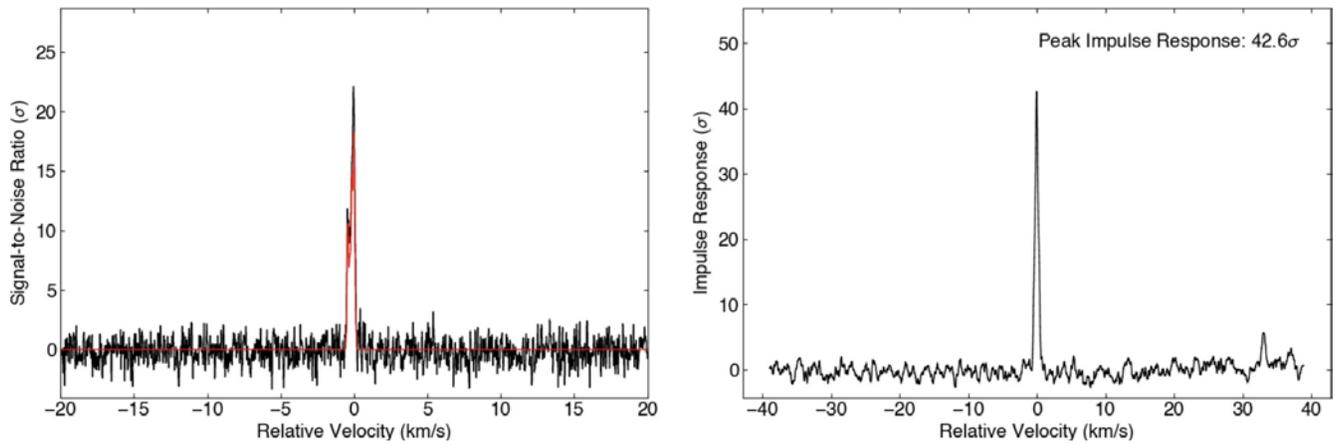

**Figure A2.** Velocity stacked spectra and matched filter response of HCCCHO toward TMC-1. Left: the line profile of propynal determined from the MCMC analysis (red) overplotted with the velocity stacked spectrum from the GOTHAM data (black). Right: Matched filter impulse response of the stacked spectrum, determined by cross-correlating the observed and simulated velocity stacks.